\documentclass[a4paper,12pt]{article}
\usepackage{bm, graphicx}
\title{Surface and electronic structure of MOCVD-grown
Ga$_{0.92}$In$_{0.08}$N investigated by UV and X-ray photoelectron
spectroscopies}

\author{B.J. Kowalski${^1}$, \and I.A. Kowalik${^1}$, \and R.J. Iwanowski${^1}$, \and J. Sadowski${^{2,1}}$, \and J.
Kanski${^2}$, \and  B.A. Orlowski${^1}$, \and J. Ghijsen${^3}$,
\and F. Mirabella${^3}$, \and E. Lusakowska${^1}$, \and P.
Perlin${^4}$, \and S. Porowski${^4}$, \and I. Grzegory${^4}$, \and
M. Leszczynski${^4}$}

\begin{document}

\maketitle

\noindent ${^1}$Institute of Physics, Polish Academy of Sciences,
Aleja~Lotnik\'{o}w~32/46, PL-02~668~Warsaw, Poland

\noindent ${^2}$Department of Physics, Chalmers University of
Technology and G\"{o}teborg University, S-412 96 G\"{o}teborg,
Sweden

\noindent ${^3}$Facult\'{e}s Universitaires Notre-Dame de la Paix,
LISE, 61 rue de Bruxelles, B-5000 Namur, Belgium

\noindent ${^4}$High Pressure Research Center, Polish Academy of
Sciences, Sokolowska~29, PL-01~141~Warsaw, Poland

\begin{abstract}
The surface and electronic structure of MOCVD--grown layers of
Ga$_{0.92}$In$_{0.08}$N have been investigated by means of
photoemission. Stability of chemical composition of the surface
subjected to Ar$^{ + }$ ion sputtering was proven by means of
X-ray photoemission spectroscopy. The analysis of the relative
intensities of In 3d, Ga 3p, and N 1s peaks showed that argon ion
bombardment does not change significantly the relative contents of
the layer constituents. Simultaneous efficient removal of the main
contaminants (O and C) was observed during the sputtering
procedure, proving that argon sputtering can be used as method for
preparation of clean Ga$_{1 - x}$In$_{x}$N surfaces.

For a clean (0001)-(1x1) surface prepared by repeated cycles of
Ar$^{ + }$ ion sputtering and annealing, electronic structure was
investigated. The band structure was explored along the $\Gamma
$-A direction of the Brillouin zone, measuring angle-resolved
photoemission spectra along the surface normal. A similar set of
data was also acquired for the same surface of GaN layer.
Comparison of the collected data revealed an additional feature at
the valence band edge, which can be ascribed to the presence of In
in the layer.

\end{abstract}



\section{Introduction}

Epilayers of Ga$_{1 - x}$In$_{x}$N are indispensable elements of
optoelectronic devices operating within green-blue range of
electromagnetic radiation spectrum \cite{nakam95,nakam97}.
Introducing indium into the GaN matrix leads to a narrowing of the
energy gap of the material from 3.5 to 1.9 eV and makes it
possible to tune the energy of light emitted by the device. The
perspective of promising applications together with interesting
physical phenomena governing the radiative recombination
efficiency have stimulated intensive investigations of this
material. Detailed information about the electronic structure is
important for both applications and research of elementary
excitations in it. To get information about electronic band
structure we must use experimental methods which enable direct
probing of electronic states. Despite its surface sensitivity
photoemission is widely accepted as one of the most suitable tools
for such studies. However, atomically clean, ordered surfaces with
unmodified chemical composition are a precondition for meaningful
photoemission experiments. Since Ga$_{1 - x}$In$_{x}$N is only
available in the form of epitaxial layers grown by molecular beam
epitaxy (MBE) or metal organic chemical vapor deposition (MOCVD),
surfaces for photoemission studies cannot be obtained by cleavage
under UHV conditions. Instead, UHV-compatible methods consisting
in Ar$^{+}$ ion sputtering and annealing have to be considered.
They have been applied in various combinations in order to prepare
surfaces of GaN epilayers for electronic structure studies
\cite{hunt93,ding96,dhesi97,beach99}. Such procedures led to
clean, well ordered (0001)-(1x1) surfaces, suitable for band
structure studies by means of angle-resolved photoelectron
spectroscopy. However, it was found for indium containing nitride
layers that the presence of In atoms modifies surface structure
(for the (0001) orientation, particularly) \cite{chen00a}, by
formation of indium-rich surface nanostructures \cite{chen00b} or
by inducing non-uniform depth distribution of constituent elements
\cite{pere01a,pere01b}. Moreover, surface sputtering of
multicomponent solid may lead to a substantial change in chemical
composition of the subsurface layer. The extent of that effect has
to be evaluated experimentally before studies of the band
structure are undertaken. Due to such obstacles, experimental
evidence concerning band structure of Ga$_{1 - x}$In$_{x}$N is
scarce, except of that referred to the edges of the bands forming
the fundamental energy gap, which were obtained mainly by various
optical measurements
\cite{osamura75,wetzel98,cluskey98,walle99,parker99,donnel01,perlin01}.
The bowing of band gap vs composition dependence in strained and
relaxed layers of Ga$_{1 - x}$In$_{x}$N was one of the main issues
in those papers. Conclusions derived from the experiments were
supported by theoretical calculations
\cite{cluskey98,walle99,wright95,pugh99,teles01}. The density of
states in the whole valence and conduction bands was investigated
by soft X-ray emission and absorption \cite{ryan02}. Changes of
the band structure as a function of indium content were reported,
in particular for the top of the valence band, in the alloys with
a degree of substitution lower than 0.1. The conclusions from that
study will be taken into account in the analysis of our
photoemission results. Some results of photoemission experiments
on InN have also been reported \cite{guo98}.

In this paper, we report photoemission investigations of
properties of MOCVD-grown Ga$_{1 - x}$In$_{x}$N. The first part
presents the X-ray photoelectron spectroscopy (XPS) study of this
material subjected to Ar$^{ + }$ ion sputtering. This was aimed at
revealing the effect of such surface treatment on chemical
composition of the surface region of the sample in order to check
if Ar$^{ + }$ sputter cleaning could be applied to surfaces
intended for photoemission measurements. It has been proved here
that such a procedure could be adequate for surface preparation of
Ga$_{1 - x}$In$_{x}$N before photoemission experiments.

The second part is devoted to the angle-resolved photoemission
investigation of the band structure of Ga$_{1 - x}$In$_{x}$N. The
sample surface was prepared by Ar$^{ + }$ ion sputtering, followed
by annealing at 600$^{o}$C. Normal emission spectra acquired for
photon energies from 26 to 76 eV enabled us to map the valence
band structure along the $\Gamma $-A direction of the Brillouin
zone. The similar set of data was collected for the corresponding
surface of a GaN layer thus enabling us to reveal an additional
feature at the valence band edge of Ga$_{1 - x}$In$_{x}$N.
Comparison of our results with those derived from theoretical
calculations of Ga$_{1 - x}$In$_{x}$N valence band density of
states \cite{perlin01} supports the assignment of the above
feature to the presence of indium ions in the studied system.

\section{Experimental}

The investigated samples were prepared in the High Pressure
Research Centre, Polish Academy of Sciences in Warsaw, Poland. The
single crystalline layers of GaN and Ga$_{1 - x}$In$_{x}$N (x =
0.08) were grown epitaxially by the MOCVD technique on the (0001)
faces of bulk GaN crystals \cite{leszcz01}. The substrate surface
was covered with a 0.5 $\mu$m thick buffer layer of GaN ---
subsequently, the layer of Ga$_{0.92}$In$_{0.08}$N with the
thickness of 100 nm was grown. The whole structure was terminated
by (0001) (Ga-polar) face. The lattice parameters (as determined
by XRD) were {\em c}=5.157 \accent23A for GaN and {\em c}=5.253
\accent23A for Ga$_{1 - x}$In$_{x}$N, with {\em a}=3.175
\accent23A for both materials. Bulk GaN crystals (the substrates)
with hexagonal crystalline structure were grown by means of high
pressure technique \cite{grzeg95}. Their dislocation densities
were as low as 10 - 100~cm$^{ - 2}$ and they had no inversion
domains on surfaces.

The influence of Ar$^{ + }$ bombardment on the Ga$_{1 -
x}$In$_{x}$N surface was studied by means of XPS. These
experiments were performed using the photoemission spectrometer
ESCA-300 at LISE FUNDP in Namur, Belgium. The base pressure in the
system was p=1$\times $10$^{ - 10}$ Torr. The photoemission was
excited by a monochromatized Al K$\alpha$ X-ray beam
($h\nu$=1486.6 eV). The total energy resolution was about 0.35 eV.
The sputtering was carried out under static Ar pressure of
5$\times $10$^{ - 7 }$ hPa and 600 V acceleration potential.

The angle-resolved (AR) photoemission experiments were carried out
at beamline 41 of the MAX-lab synchrotron radiation laboratory in
Lund, Sweden. The overall energy resolution was kept around 150
meV, and the angular resolution was about 2$^{o}$. The angle
between incoming photon beam and the surface normal was kept at
45$^{o}$. The surface crystallinity was assessed by low energy
electron diffraction (LEED) which indicated a hexagonal (1x1)
symmetry.

\section{Results and discussion}

\subsection{Core level spectroscopy}

\begin{figure}
\begin{center}
\includegraphics[height=14cm]{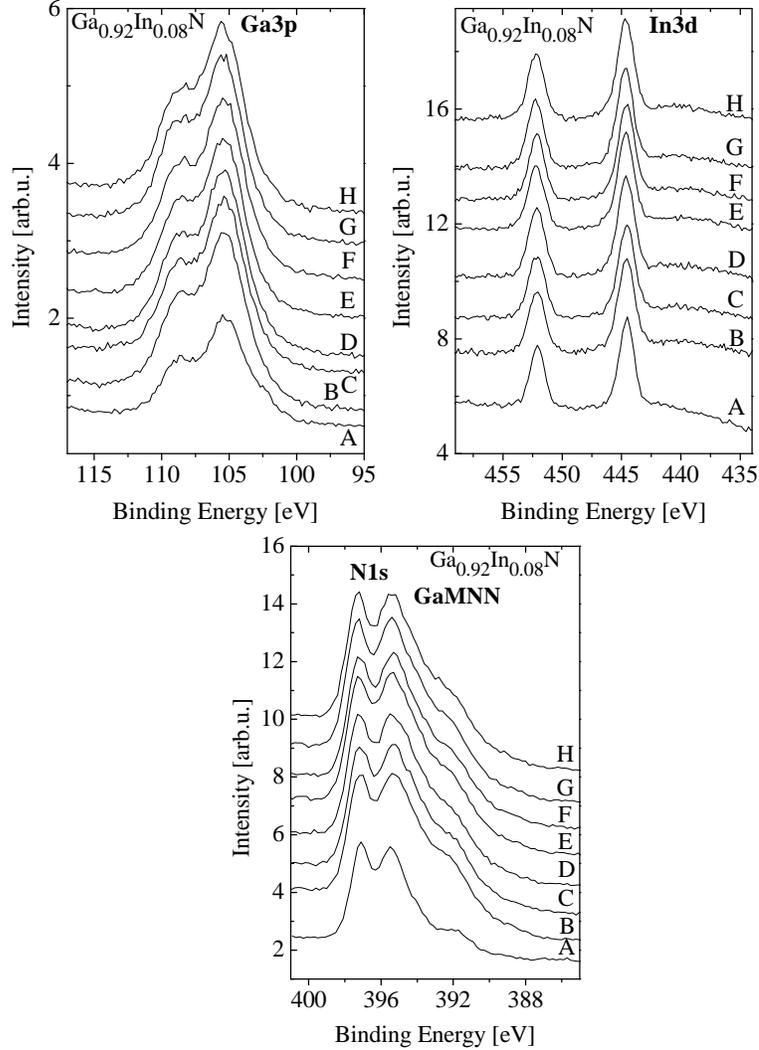}
\end{center}
\caption{\label{Gainn}XPS spectra of Ga 3p, In 3d and N 1s
recorded after subsequent stages of Ar$^{ + }$ sputtering. Curves
marked A correspond to the "as-mounted" surface. Spectra B-H were
obtained after 45, 90, 150, 210, 270, 330 and 390 min. of Ar$^{+}$
sputtering.}
\end{figure}

X-ray photoemission spectroscopy was used for monitoring the
process of surface cleaning of Ga$_{1 - x}$In$_{x}$N epilayer,
particularly, the changes of the chemical composition of the
surface under sputtering. The core-level spectra of the Ga$_{1 -
x}$In$_{x}$N components (In 3d, Ga 3p, N 1s) and of the main
surface contaminants (O 1s and C 1s) were acquired, after
sequential sputtering steps. Fig.~\ref{Gainn} shows the spectra of
core levels of sample constituents. XPS binding energies were
referenced to the C 1s peak of adventitious carbon at 284.8 eV
\cite{moulder94}.

For In 3d and Ga 3p, the maxima are well discernible and
spin--orbit splitting has been resolved. The N 1s peak partially
overlaps the Auger GaMNN feature. The overall shape of these
spectra did not change during sputtering. No appearance or
disappearance of multiple components of the features could be
discerned.

\begin{figure}
\begin{center}
\includegraphics[height=10.2cm]{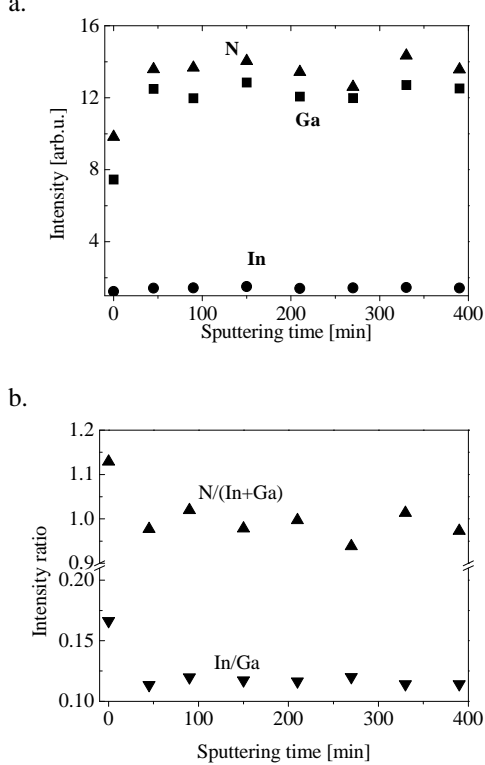}
\end{center}
\caption{\label{trends}Changes of the intensities and intensity
ratios of core level maxima, occurring during Ar$^{+}$ sputtering
of Ga$_{0.92}$In$_{0.08}$N samples.}
\end{figure}

In order to evaluate possible changes in chemical composition of
the surface and subsurface layer of the sample the peak
intensities were determined as integrated area under the
corresponding peaks. For N 1s, the area was measured under a
single Gaussian extracted from the feature consisting of N 1s and
GaMNN. The intensities were normalized to experimental conditions
(number of scans, acquisition time etc.) and atomic sensitivity
factors. For all three core levels we observed an increase of
intensity after the first stage of sputtering (Fig.
~\ref{trends}a). It was strongest for Ga 3d and relatively weak
for In. Further sputtering did not lead to intensity changes,
within the accuracy of our analysis. As a consequence, the
intensity ratio In/Ga decreased from 0.17 (prior to sputtering) to
the saturation level of 0.11 after 45 min. sputtering time
(Fig.~\ref{trends}b). The latter value corresponds to the
composition parameter x equal to 0.1 and remains reasonably
consistent with its real value of 0.08. The observed change could
be related to the reported non-uniform depth-compositional
distribution in Ga$_{1 - x}$In$_{x}$N layers grown by MOCVD
\cite{pere01a,pere01b}. The N/(In+Ga) ratio also decreased after
the first sputtering, most probably due to partial deficiency of
nitrogen at the surface (induced by sputtering). Generally, this
ratio is reasonably close to expected value of 1. A surface
contribution of metal atoms, appearing due to sputtering, could
easily be observed for Ga 3d at photon energies of 40-60 eV for
$(000\overline 1)$ \cite{kowalski03} as well as (0001) surface.
Such a component could not be discerned in the reported XPS
spectra. It can be attributed to lower energy resolution and
markedly higher electron escape depth for X-rays.

The measurement and analysis of the O 1s and C 1s spectra showed
that Ar$^{+}$ ion sputtering efficiently removes these elements
from the sample surface. The most significant decrease of their
content was observed after the first stage of sputtering (i.e.
after 45 min. of Ar$^{+}$ exposure). For the "as mounted" surface,
the O 1s feature could be decomposed into two maxima centered at
531.2 and 533.4 eV. According to Watkins {\em et al.}
\cite{watkins99} the O 1s peak component at higher BE correspond
to oxygen atoms adsorbed at the surface of GaN; the other one ---
to oxygen bound in the bulk of the crystal. We found that Ar$^{+}$
ion cleaning leaves a trace amount of oxygen and carbon in the
surface layer probed by XPS. This coincides with the results of
Prabhakaran {\em et al.} \cite{prabhakaran96} for surface of GaN.
They attributed the residual C 1s emission to carbon atoms
incorporated into the epilayer during its growth.

Concluding the analysis of cleaning the layer by Ar$^{+}$
sputtering, one has to note that surface contaminants are removed
efficiently and the energy positions of the core level peaks
change toward lower BE due to removal of oxidized surface layer.
However, the shape of Ga 3p, In 3d and N 1s peaks do not change
during sputtering (no oxide-related components can be
unequivocally revealed).

The surface morphology of the investigated samples was also
studied by atomic force microscopy (AFM). A comparison of the
patterns obtained before and after sputtering showed that such a
process did not lead to marked destruction of the surface or to
increase of its roughness. However, structures with step height of
1 or 1.5 nm and horizontal width of about 300 nm became
discernible on the sputtered surface. The step height corresponds
roughly to 2-3 lattice parameters along the {\em c} axis.

In conclusion we can state that Ar$^{+}$ ion bombardment does not
change the chemical composition of the surface and subsurface
region of Ga$_{1 - x}$In$_{x}$N layers markedly. Simultaneously,
the sputtering provides significant reduction of oxygen and carbon
concentration down to the level, where they do not influence the
core--level spectra of Ga, In and N (see Figs.~\ref{Gainn}). Thus,
X-ray photoemission spectroscopy shows that Ar$^{+}$ ion
bombardment can be used as a UHV-compatible method for preparation
of clean Ga$_{1 - x}$In$_{x}$N surfaces for photoemission
experiments.

\subsection{Angle-resolved photoelectron spectroscopy}

Taking into account the results reported in the previous
subsection we undertook an attempt to study band structure of
Ga$_{1 - x}$In$_{x}$N epitaxial layer cleaned in situ by multiple
cycles of Ar$^{+}$ ion sputtering and annealing at 600$^{o}$C
under UHV conditions. The most important aim of the study was
assessment of In-related changes in the band structure of the
solid solution. Therefore, parallel measurements were carried out
for samples of Ga$_{0.92}$In$_{0.08}$N and GaN. The samples were
fixed on the sample holder next to each other and cleaned
together. So, the surface preparation, as well as the experimental
conditions, were identical for both samples.

\begin{figure}
\begin{center}
\begin{tabular}{lc}
\includegraphics[height=4.5cm]{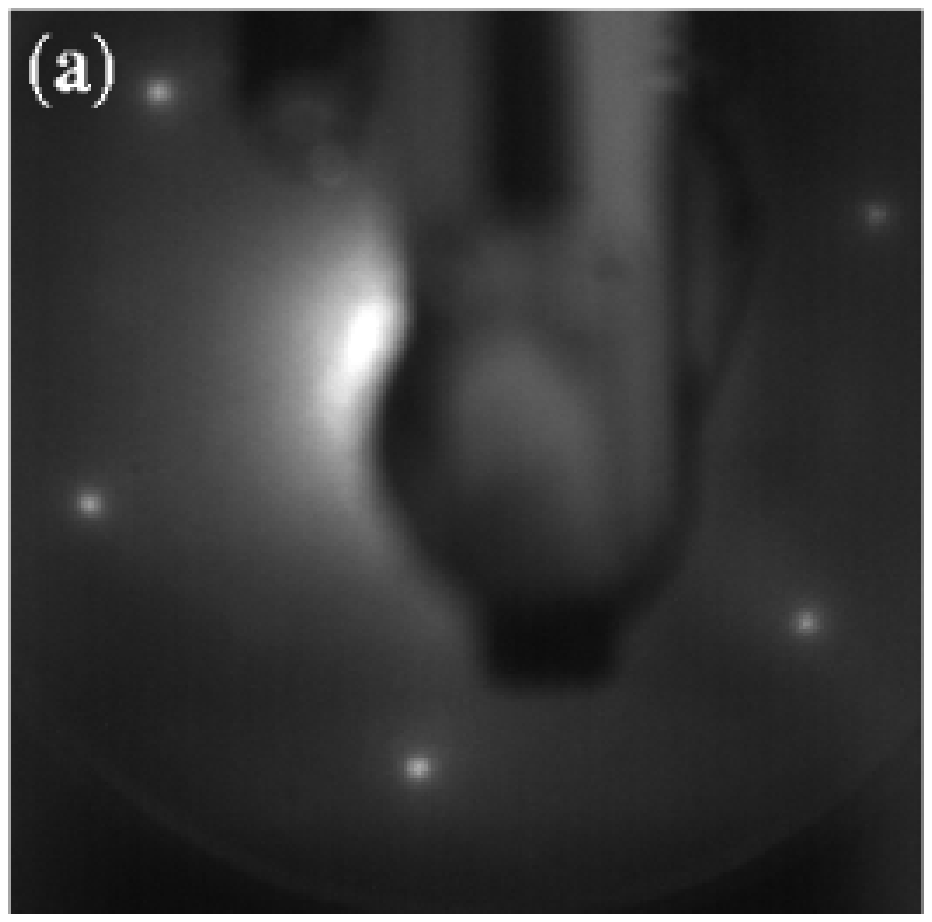}\\
\includegraphics[height=4.5cm]{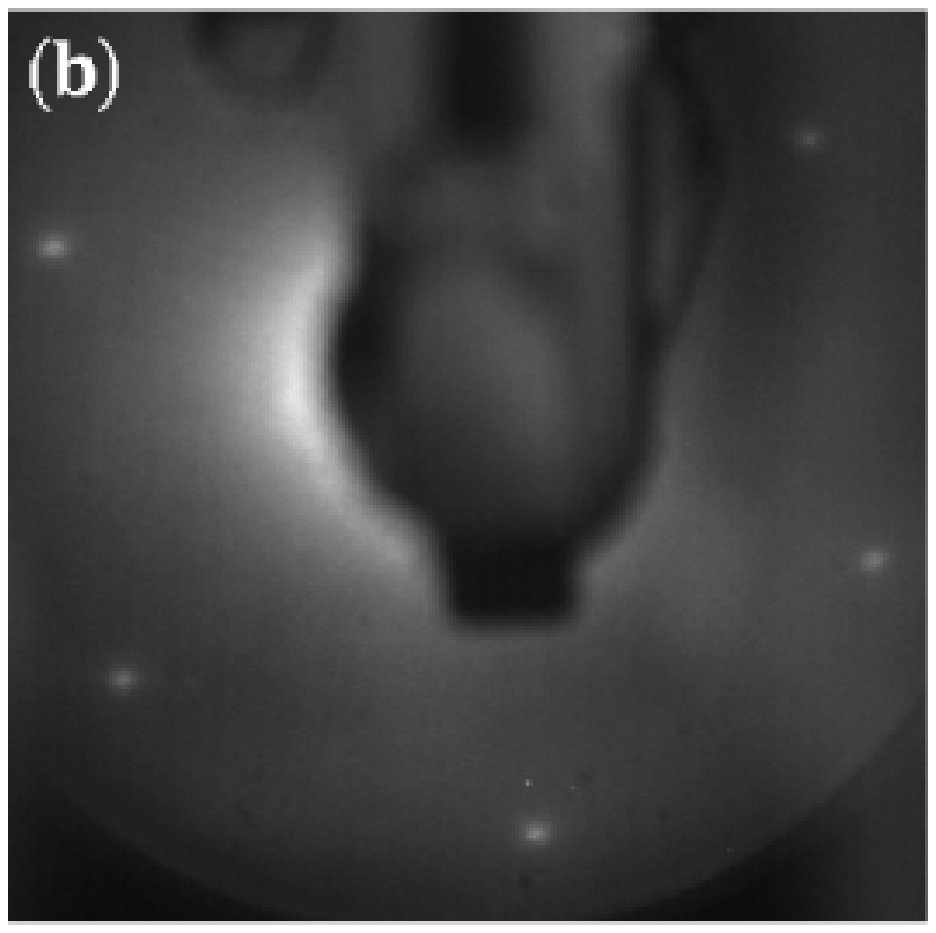}\\
\end{tabular}
\end{center}
\caption{\label{leed}The LEED patterns obtained for the electron
energy of 94.9 eV from the GaN$(000\overline 1)$ surface {\em in
situ} sputtered with Ar$^{ + }$ ions and annealed at 600$^{o}$C
(a) and for the electron energy of 93.9 eV from the corresponding
surface of Ga$_{0.92}$In$_{0.08}$N prepared by the same method
(b).}
\end{figure}

The crystallinity of the surfaces was checked using low energy
electron diffraction (LEED) method. The LEED patterns indicated a
hexagonal (1x1) symmetry of the surface for GaN as well as for
Ga$_{1 - x}$In$_{x}$N (Fig.~\ref{leed}). The surface composition
of both samples was checked during the preparation process by
spectroscopy of the Ga~3d level. Results of previous investigation
concerning preparation of GaN(0001) (Ga-polar) surface by Ar$^{ +
}$ ion bombardment and annealing showed that any changes in the
surface conditions clearly manifested themselves in shape and
position of Ga 3d structure. Sputtering efficiently removed
oxidized Ga --- it resulted in disappearance of relevant component
of Ga 3d feature (observed for "as grown" sample). Hence, after
sputtering the Ga~3d feature was centered at the BE corresponding
to pure gallium. Annealing of the sputtered sample resulted in
decrease of the contribution related to pure gallium and
appearance of the peak of Ga bound to N. These changes were
clearly correlated with sharpening of the LEED pattern. Such a
preparation procedure resulted in the clean, ordered
GaN(0001)-(1x1) surface suitable for angle-resolved photoemission
study \cite{plucinski02}. Thus, the progress in the cleaning the
surface during the sample treatment in this work was monitored by
the observation of similar changes in the Ga 3d feature.

\begin{figure}
\begin{center}
\includegraphics[height=8.6cm]{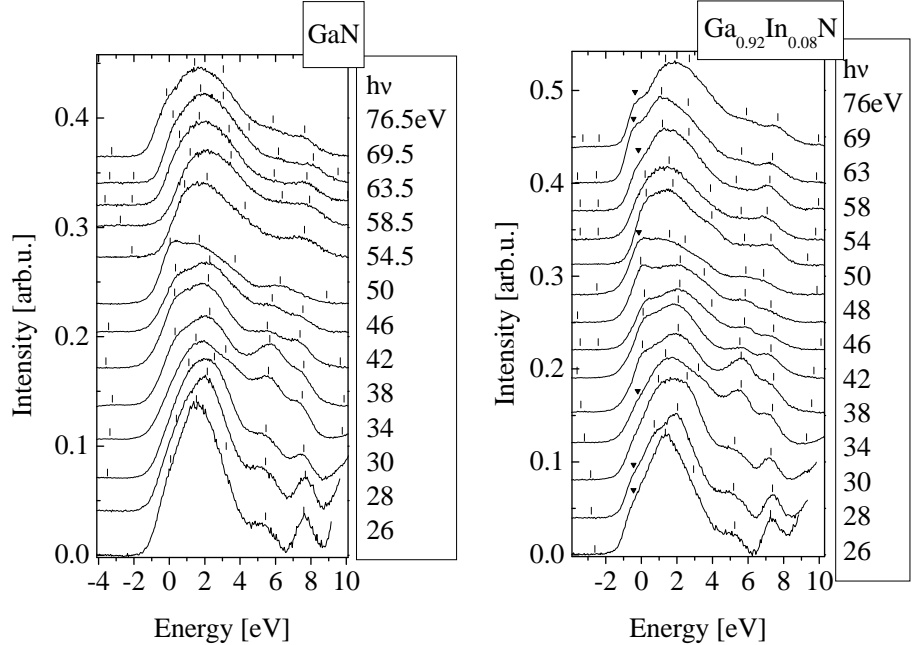}
\end{center}
\caption{\label{edcnorm}Angle-resolved photoemission spectra taken
at normal emission from the GaN~(0001) and
Ga$_{0.92}$In$_{0.08}$N~(0001) surfaces. Ticks show the features
discerned in the spectra and used to draw the experimental band
structure diagram. Triangles mark features corresponding to
emission from In-related states at the valence band maximum (see
Fig.~\ref{bands}).}
\end{figure}

\begin{figure}
\begin{center}
\includegraphics[height=9cm]{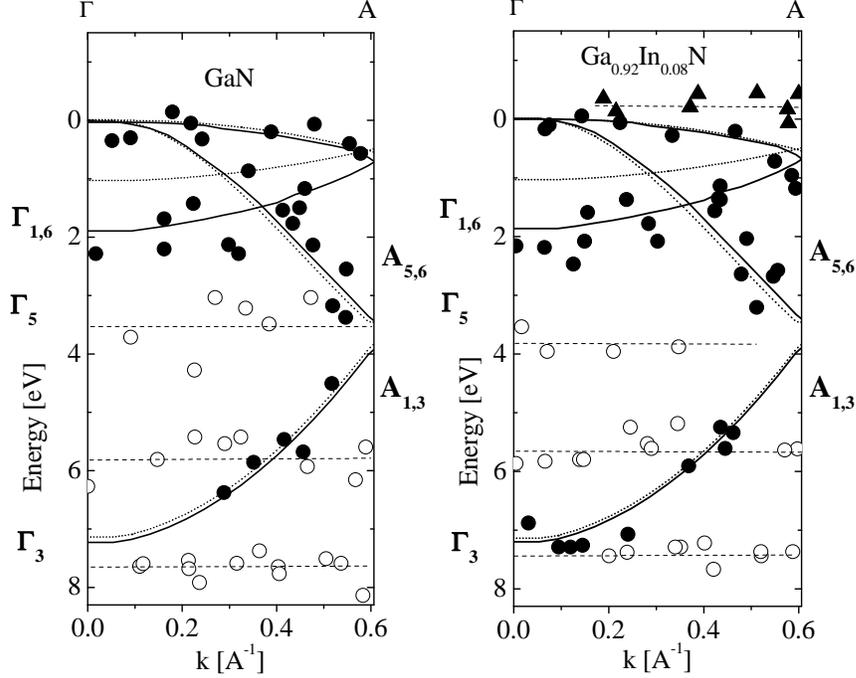}
\end{center}
\caption{\label{bands}The experimental band structure diagrams
along the $\Gamma$ -A direction derived from the data shown in
Fig.~\ref{edcnorm} for GaN and Ga$_{0.92}$In$_{0.08}$N. All
symbols mark the experimental points. Experimental points, which
can be ascribed to emission from the bulk band structure, within a
free-electron final state model, are marked with full dots. Full
lines have been drawn through the points corresponding to the bulk
bands (dotted lines show the band structure calculated by Vogel
{\em et al.} \cite{vogel97}). Horizontal dashed lines and open
dots mark dispersionless features, possibly occurring due to
surface--related states or due to "non-direct" transitions from
high density--of--states regions. Full triangles show a feature
brought about by admixture of In. The origin of the energy axis
for GaN was set at the valence band maximum. For easy comparison
the same scale of energy was used for Ga$_{0.92}$In$_{0.08}$N.}
\end{figure}

Angle-resolved photoemission was applied to study the electronic
band structure along the $\Gamma$-A direction in the Brillouin
zone. Making use of synchrotron radiation we were able to measure
the photoemission spectra for the energy range of 26-80 eV. Sets
of normal emission spectra were acquired for
Ga$_{0.92}$In$_{0.08}$N(0001)-(1x1) and for the corresponding
surface of GaN (Fig.~\ref{edcnorm}). A Shirley-type background was
subtracted from the spectra. The origin of the BE scale has been
adjusted to the valence band maximum (as determined for GaN from
the experimental band structure diagram (Fig.~\ref{bands})). The
Fermi level position in the samples was around the edge of the
conduction band, as estimated by photoemission measurement for a
metallic reference sample in electric contact with investigated
layers. This justifies excluding relation between features
revealed at the valence band edge of Ga$_{0.92}$In$_{0.08}$N and
appearance of any metal layer or islands on the sample surface.

The overall shape of the spectra seems reasonably similar to those
previously reported for GaN (0001) surfaces
\cite{ding96,dhesi97,maruyama99}, taking into account differences
between results obtained by various groups. Certain discrepancies
should be attributed to modifications of surface preparation
procedures. Irrespectively of differences in relative heights of
the features, the spectra consist of two main groups of maxima.
The first one, ranging from the valence band maximum to the BE of
about 4 eV corresponds mainly to emission from the top valence
bands. The second part of the curves is dominated by two features
with weak dispersion (at BE of 5-6 eV and about 8 eV) and emission
from the deeper valence band \cite{strasser99}. The corresponding
spectra collected for GaN and Ga$_{0.92}$In$_{0.08}$N are
generally similar. However, a clearly discernible feature appears
at the leading edge of the spectra of In-containing layer for
higher photon energies (58--76 eV). This feature overlaps with a
weaker shoulder appearing for GaN at slightly higher binding
energies. A similar, although weaker, modification can also be
resolved for the spectra recorded for the lowest photon energies.

Two weak features could be revealed for the energies above the
valence band maximum --- at -2.0 and -3.4 eV for GaN, at -2.4 and
-3.4 eV for Ga$_{0.92}$In$_{0.08}$N. It was shown by Wang {\em et
al.} \cite{wang01} that electronic states occur in the energy gap
on the (0001)-(1x1) surface. However, it is difficult to relate
our findings with the results of their calculation due to well
known difficulty in determining of energy gap by DFT-LDA
calculations. The energy position of the second feature coincides
well with the width of GaN energy gap. The presence of such
emission, although weak, may suggest strong band bending downwards
and formation of an inversion layer at the surface of the samples.
These features need and seem to deserve more detailed
investigation.

In order to gain insight into details of the band structure of
both systems the sets of photoemission spectra measured for normal
emission were transformed into a band structure diagram along the
$\Gamma$ -A direction (Fig.~\ref{bands}). Features appearing in
the spectra were revealed and their energy positions were
determined by means of an inverted second derivative method
(marked with ticks and triangles in Fig.~\ref{edcnorm}). Those BE
values were used as the input data for calculation of
corresponding k-vectors. We applied the free-electron final-state
model using the formula:

\begin{equation}\label{first}
k_{\perp}=\sqrt{\frac{2m}{\hbar ^2}(E_{kin}+V_{0})}-G_{\perp}
\end{equation}

The value for inner potential of the crystal V$_{0}$=15.5 eV was
determined by analysis of the curvature of the experimental bands
and their correspondence to the available calculated band
structure parameters of GaN. For almost all points, $G_{\perp}$
was equal to three basis vectors of the reciprocal lattice. In
Fig.~\ref{bands} points derived from the experimental data are
marked with full and open symbols. The full lines have been drawn
through the data points corresponding to the bulk bands (full
dots) while dotted lines show the band structure calculated by
Vogel {\em et al.} \cite{vogel97}. Dashed lines (drawn through
open symbols) mark dispersionless features, possibly occurring due
to emission from surface-related states or "non-direct" emission
from a high density--of--states parts of the bands.

The width of the top part of the valence band (the energy
separation between $\Gamma _{1,6}$ and $\Gamma _{5}$) is markedly
larger in the experimental band structure than in the calculated
one (about 1.9 and 1.0 eV, respectively). Similar discrepancy was
previously observed for the $(000 \overline 1)$ surface of bulk
GaN crystals \cite{kowalski03}, although the difference was
smaller. The total width of the valence band at the $\Gamma$ point
($\Gamma _{1,6}-\Gamma _{3}$) is consistent with the results of
calculations.

\begin{figure}
\begin{center}
\includegraphics[height=5cm]{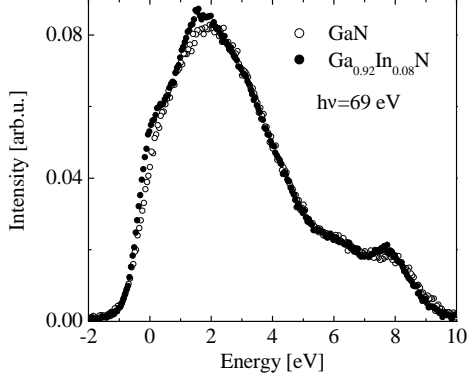}
\end{center}
\caption{\label{Comp} Photoemission spectra of GaN and
Ga$_{0.92}$In$_{0.08}$N (taken for $h\nu$= 69 eV) superposed in
order to visualize the differences in the valence band density of
states.}
\end{figure}

A comparison of the experimental band diagrams of both materials
shows that the main modification of the band structure brought
about by the admixture of In is an additional feature at the
energy of 0.2 eV (marked with solid triangles). It led to an
upward shift of the valence band edge of Ga$_{0.92}$In$_{0.08}$N.
A similar effect was observed by X-ray absorption \cite{ryan02}. A
shift of the valence band maximum by 0.15 eV was observed in N 2p
soft x-ray emission spectra between GaN and Ga$_{0.9}$In$_{0.1}$N.
Ryan {\em et al.} found it in agreement with results of empirical
pseudopotential calculations of In wave functions and energy gap
behavior in low-In content Ga$_{1 - x}$In$_{x}$N. Bellaiche {\em
et al.} \cite{bellaiche99} showed that the hole wave function
localized around In could be interpreted as a combination of
hybridized In functions with those of GaN valence band. Such a
localization of states occurred above as well as below the valence
band maximum of GaN. Our observation is also consistent with the
results of first-principles calculations of band structure
performed for GaN and Ga$_{1 - x}$In$_{x}$N \cite{perlin01}. The
calculations showed that some p states of nitrogen are pushed up
from the top of the valence band, due to hybridization of In
states and states of GaN. Thus, it seems to be justified to
ascribe the dispersionless feature revealed in the Ga$_{1 -
x}$In$_{x}$N band structure to In(p,d)-N(p) hybridization states.
It can easily be discerned in Fig.~\ref{Comp}, where the
differences between the spectra recorded for $h\nu$= 69 eV were
visualized by direct comparison. The difference at the energy of
1.5 eV overlaps with bulk valence bands and cannot be discerned in
the experimental band structure diagrams.

Dispersionless features discernible in the experimental band
structures of GaN and Ga$_{1 - x}$In$_{x}$N can be related to
surface states or to non-direct emission from high
density-of-states regions in the Brillouin zone. According to Wang
{\em et al.} \cite{wang01}, a surface band occurs just below the
lowest valence band. Its energy at the $\Gamma$ point is equal to
8 eV (with respect to the valence band maximum). Such a surface
band is characteristic of the GaN(0001)-(1x1) surface terminated
by additional layer of Ga atoms bound at  $T _{4}$ positions.
Emission from that state may contribute to the feature found at
about 8 eV for both materials. However, one has to take into
account the possibility of "non-direct" emission from the lowest
part of the valence band. Similarly, the feature at 4 eV can be
connected with a surface band occurring around the $\Gamma$ point
on a clean Ga-terminated (0001) surface, but again emission from
high density of states region may be important. For the feature at
6 eV, we can consider a surface band between the K and M points of
the surface Brillouin zone \cite{wang01} and emission from bulk
band at the K point.

\section{Summary}

The formation of Ga$_{1 - x}$In$_{x}$N(0001) surface by Ar ion
sputtering was investigated by X-ray photoelectron spectroscopy
and atomic force microscopy. The intensities of the peaks related
to emission from the Ga$_{1 - x}$In$_{x}$N layer constituent
elements (Ga, In, N) were measured for the sample "as-mounted" and
after subsequent stages of sputtering. For sputtering with Ar$^{ +
}$ ions having relatively low kinetic energy (600 eV) no
substantial modification of In/Ga ratio was observed, except for a
weak change in chemical composition at the initial stage of
sputtering. The main contaminants (oxygen and carbon) were
efficiently removed from the surface. Therefore, we believe that
argon ion sputtering can be used as a UHV-compatible method for
preparation of clean Ga$_{1 - x}$In$_{x}$N surfaces for
photoemission experiments.

Comparative studies of the (0001)-(1x1) surfaces of
Ga$_{0.92}$In$_{0.08}$N and GaN layers have been performed by
angle-resolved photoemission in order to derive their band
structure. The experimental band structures have been mapped along
the $\Gamma$-A direction of the Brillouin zone.

An additional dispersionless feature at the valence band maximum
has been observed for Ga$_{0.92}$In$_{0.08}$N. The analysis based
on comparison with available results of theoretical calculations
enabled us to relate this feature to the states pushed up from the
top of the valence band, due to hybridization of In and N states.

{\bf Acknowledgments} This work was supported by KBN (Poland)
project 2P03B 046 19 and by the European Commission - Access to
Research Infrastructure action of the Improving Human Potential
Programme (realized by MAX-lab facility in Lund). J.G. is
supported by NFSR (Belgium) and F.M. by FRIA.

\end{document}